\documentclass[graybox]{svmult}

\usepackage{mathptmx}       % selects Times Roman as basic font
\usepackage{helvet}         % selects Helvetica as sans-serif font
\usepackage{courier}        % selects Courier as typewriter font
\usepackage{type1cm}        % activate if the above 3 fonts are
\usepackage{makeidx}         % allows index generation
\usepackage{graphicx}        % standard LaTeX graphics tool
\usepackage{multicol}        % used for the two-column index
\usepackage[bottom]{footmisc}% places footnotes at page bottom

\makeindex             % used for the subject index
\begin{document}

\title*{Stellar driven evolution of hydrogen-dominated atmospheres of terrestrial exoplanets}
\author{K.G.~Kislyakova, M.~Holmstr{\"o}m, H.~Lammer, and N.V.~Erkaev}
\institute{K.G.~Kislyakova \at Space Research Institute, Austrian Academy of Sciences, Graz, A-8042, Austria, \email{kristina.kislyakova@oeaw.ac.at}
\and M.~Holmstr{\"o}m \at Swedish Institute of Space Physics, P.O. Box 812, SE-98128 Kiruna, Sweden, \email{matsh@irf.se} 
\and H.~Lammer \at Space Research Institute, Austrian Academy of Sciences, Graz, A-8042, Austria, \email{helmut.lammer@oeaw.ac.at}}
\maketitle

\abstract{In the present chapter we present the results of evolutionary studies of exoplanetary atmospheres. We mostly focus on the sub- to super-Earth domain, although these methods are applicable to all types of exoplanets. We consider both thermal and nonthermal loss processes. The type of thermal loss mechanism depends on so-called escape parameter $\beta$, which is the ratio of the gravitational energy of a particle to its thermal energy. While $\beta$ is decreasing, an exoplanet switches from classical Jeans to modified Jeans and finally to blow-off escape mechanisms. During blow-off the majority of the atmospheric particles dispose of enough energy to escape the planet's gravity field. This leads to extreme gas losses.\newline
Although nonthermal losses never exceed blow-off escape, they are of significant importance for planets with relatively weak Jeans-type escape. From the diversity of nonthermal escape mechanisms, in the present chapter we focus on ion pickup and discuss the importance of other loss mechanisms. 
The general conclusion of the chapter is, that escape processes strongly shape the evolution of the exoplanets and determine, if the planet loses its atmosphere due to erosion processes or, on the contrary, stays as mini-Neptune type body, which can probably not be considered as a potential habitat as we know it.
}

%\abstract{}

\section{Introduction }
\label{sec:intro}

One of the most important questions still awaiting the solution in the present-day exoplanetary science is the determination of possible evolution scenarios of exoplanets, which could explain all currently observed types.

All exoplanets as well as planets in the Solar system experience atmospheric mass losses. There exist a number of observations and measurements of losses from Solar system terrestrial planets (see, for example, \cite{Lundin2011}), and observations in Ly-$\alpha$ line for Hot Jupiters HD209458b and HD189733 \cite{Vidal-Madjar2003, Lecavelier2010}. Excess absorption in Ly-$\alpha$ is usually explained as observational evidence for mass loss \cite{Bourrier2013}.

Here we focus on moderate size sub- to super-Earth type planets (we call them terrestrial in the present chapter), i.e. we exclude massive exoplanets on close and remote orbits from consideration. According to present-day modeling, terrestrial-sized planets should be common in the Universe \cite{Broeg2009}. We make a short review about the atmospheric evolution of these moderate size planets and factors, which define it, as it is understood at present.

The atmosphere mass and composition of a terrestrial planet is defined, first, by the formation conditions and, second, by the following stellar-driven escape process. Atmospheres may take origin in the primordial nebula, where all exoplanets are believed to be born \cite{Ikoma2006, Lammer2013a, Lammer2014}, or be later outgassed from the planetary interiors during solidification \cite{Elkins-Tanton2008}. In the second case one speaks about secondary atmosphere. Both types of atmosphere formation are believed to be important and work together on the atmosphere evolution.

Resent findings from ESO's High Accuracy Radial velocity Planetary Searcher (HARPS) project and NASA's Kepler space observatory reveal that from the radius-mass relation and the resulting density of discovered super-Earths, these bodies probably have rocky cores, but are surrounded by significant H/He, H$_2$O envelopes, or both (e.g., http://www.exoplanet.eu). This findings demonstrate, that even if an exoplanet has a terrestrial size, its evolution may still differ much from the one of the Earth leaving the planet as a mini-Neptune type body. These discoveries raise the question about different evolution scenarios these exoworlds went in comparison to the Solar system terrestrial planets and demonstrate, that not all of them could rid of their primordial dense envelopes. After the atmosphere formation is completed, its loss and evolution are defined by intensity of escape processes.

Several types of escape mechanisms from planetary atmospheres are known. The light gases like H and He are the most inclined to escape, however, heavier species can be lost too. It was shown by several authors \cite{Tian2008a, Tian2008b, Lichtenegger2010, Lammer2013a}, that nitrogen-dominated atmospheres are not easy to be kept during the first extreme phases of the star newly arrived at ZAMS due to extreme atmospheric expansion. Other species can be lost due to drag with the escaping hydrogen (this is considered as a possible explanation for Xe fractionation in the atmosphere of the Earth and was estimated, e.g., by \cite{Hunten1987}). CO$_2$ atmosphere should not experience such strong expansion \cite{Kulikov2006} and could be kept more easily, as one may see in the case of Venus. 

Roche lobe effects can play a very important role for inflated close-in exoplanets, which fill their Roche lobe and can consequently lose big amounts of mass (e.g. \cite{Erkaev2007, Lecavelier2004}). Close-in terrestrial planets lose their primordial envelopes relatively easy \cite{Leitzinger2011} and develop into hot bodies without atmosphere like Corot-7b. However, it is not so clear how mediate-sized planets in sub- to super-Earth mass and size domain can evolve in the HZs of their stars. Can they ofter evolve to Class I habitats with nitrogen atmosphere and life favorable conditions \cite{Lammer2009} which resemble the Earth, or do they usually evolve differently?

In this chapter we present the summary of our evolutionary studies for hydrogen-rich terrestrial-type exoplanets located in the HZ of its parent star. We consider thermal and nonthermal ion pickup losses and discuss their possible influence on the planetary evolution. The main aim is to estimate the conditions under which these primordial envelopes may be lost, leaving back an exoplanet with possible Earth-type atmosphere.

\section{Thermal escape}
\label{sec:th}

On the basis of the sources of the host stars energy input into the upper atmosphere, one can separate two main escape categories: thermal escape of neutral particles and non-thermal escape of neutrals and ions.
Jeans escape is the classical thermal escape mechanism based on the fact that the atmospheric particles have velocities according to the Maxwell distribution. Individual particles in the high tail of the distribution may reach escape velocity at the exobase altitude, where the mean free path is comparable to the scale height, so that they can escape from the planet's atmosphere. When the thermosphere temperature rises due to heating by the stellar XUV radiation, the bulk atmosphere starts hydrodynamically to expand with consequent adiabatic cooling. In such a case the velocity distribution at the exobase level is described by a shifted Maxwellian (e.g., \cite{Tian2008a, Erkaev2013}). This regime can be called controlled hydrodynamic escape which resamples a strong Jeans-type escape but is still weaker compared to classical blow-off, where no control mechanism influences the escaping gas. If the XUV heating continues to increase so that the ratio between the gravitational and thermal particle energy becomes $\le$1.5 than so-called blow-off occurs leading to a stronger escape in comparison to the Jeans and even the controlled hydrodynamic escape.
As it is shown by \cite{Erkaev2013}, depending on the availability of possible IR-cooling molecules and the planets average density, hydrogen-rich “super-Earths” orbiting inside the HZ will reach hydrodynamic blow-off only for XUV fluxes several 10 times higher compared to today's Sun, which is the case for the early evolution of a Solar-type star \cite{Guedel2007}. Most of their lifetime the upper atmospheres of these planets will experience non-hydrostatic conditions, but not blow-off. In such case the upper atmosphere expands hydrodynamically and the loss of the upward flowing gas results in controlled hydrodynamic escape.
The blow-off stage is more easily reached at less massive hydrogen-rich planets with mass equal to that of the Earth. These planets experience hydrodynamic blow-off for much longer, and change from the blow-off regime to the controlled hydrodynamic escape regime for XUV fluxes which are $\le$10 times of today's Sun. Because of XUV heating and expansion of their upper atmospheres, both an exo-Earth and a super-Earth should produce extended exospheres or hydrogen coronae distributed above possible magnetic obstacles defined by intrinsic or induced magnetic fields. In such case the hydrogen-rich upper atmosphere will not be protected by possible magnetospheres like on present-day Earth, but could be eroded by the stellar wind plasma flow and lost from the planet in the form of ions \cite{Khodachenko2007, Kislyakova2013}. 

In this part of the chapter are presented the calculations of the losses of captured hydrogen envelopes from protoplanets having masses in a range between sub-Earth-like bodies of 0.1 $M_\oplus$ and super-Earths with 5$M_\oplus$, assuming that their rocky cores had formed before the nebula gas dissipated. In the thermal escape calculations we focused at bodies within the habitable zone (HZ) of a G star. These results are published in \cite{Lammer2014}. This article is a continuation of research performed in \cite{Erkaev2013}, where the same methods were used to estimate the amount of gas an Earth-type planet and a super-Earth ($M=10M_\oplus$, $R=2R_\oplus$) can lose in a Habitable Zone (HZ) of a G dwarf. The same code was also used by the the team in \cite{Lammer2013b} to estimate the thermal losses from Kepler-11 super-Earths.

The model solves the system of the hydrodynamic equations for mass,
\begin{equation}
\frac{\partial \rho R^2}{\partial t} + \frac{\partial \rho v R^2}{\partial R}= 0,
\end{equation}
momentum,
\begin{eqnarray}
\frac{\partial \rho v R^2}{\partial t} + \frac{\partial \left[ R^2 (\rho v^2+P)\right]}{\partial R} =\rho g R^2 + 2P R,
\end{eqnarray}
and energy conservation
\begin{eqnarray}
\frac{\partial R^2\left[\frac{\rho v^2}{2}+\frac{P}{(\gamma-1)}\right]}{\partial t}
+\frac{\partial v R^2\left[\frac{\rho v^2}{2}+\frac{\gamma P}{(\gamma - 1)}\right]}{\partial R}=\nonumber\\
\rho v R^2 g + q_{\rm XUV} R^2.
\end{eqnarray}
The distance $R$ corresponds to the radial distance from the center of the protoplanetary core, $\rho, P, T, v$ are the mass density,pressure, temperature and velocity of the nonhydrostatic outward flowing bulk atmosphere. $\gamma$ is the polytropic index, $g$ the gravitational acceleration and $q_{\rm XUV}$ is the XUV
volume heating rate.

Table \ref{t_mnras} summarizes the results obtained by \cite{Lammer2014}. The amount of gas accreted by the planet depends on nebula properties (dust depletion factor $f_{\rm env}$ and the core mass). The calculation was performed in the domain from the inner boundary $R_0$ up to critical point $R_c$, where the Knudsen number is equal to 0.1. The table summarizes the results: total accreted atmosphere mass $M_{\rm atm}$ and total losses $L_{\rm \Delta t}$ in [\%] during the first 90 Myr after the parent G star arrived at ZAMS. Escape rates $L_{\rm th}$ in units of g$\cdot$s$^{-1}$ and total mass loss during this period were calculated as well and can be found in \cite{Lammer2014}.

\begin{table}
\renewcommand{\baselinestretch}{1}
\caption{Captured hydrogen envelopes in units of EO$_{\rm H}$ and
integrated losses $L_{\rm \Delta t}$ during 90 Myr in \%, during the G-star XUV
saturation phase with a 100 times higher XUV flux compared to that
of the present solar value for protoplanets
with core masses of 0.1$M_{\rm \oplus}$, 0.5$M_{\rm \oplus}$,
1$M_{\rm \oplus}$, 2$M_{\rm \oplus}$, 3$M_{\rm \oplus}$, and
5$M_{\rm \oplus}$. All losses are calculated for two heating
efficiencies $\eta$ of 15\% and 40\% and the three nebula
conditions  for dust depletion factors $f\sim$0.001--0.1. Losses $L_{\rm \Delta t}$ are given in [\%], atmosphere mass $M_{\rm atm}$ is given in [EO$_{\rm H}$], accretion rate in [yr$^{-1}$].}
\begin{center}
\begin{tabular}{cccccccc}
     &                          &$f=0.001$,$\eta$=15-40\%   &    &    $f=0.01$,$\eta$=15-40\%         &      & $f=0.1$,$\eta$=15-40\%        \\\hline

$M_{\rm core}$ &  $\frac{\dot{M}_{\rm acc}}{M_{\rm pl}}$ & $M_{\rm atm}$ & $L_{\rm \Delta t}$ &  $M_{\rm atm}$  &  $L_{\rm \Delta t}$ &  $M_{\rm atm}$  & $L_{\rm \Delta t}$           \\\hline\hline
0.1  &              $10^{-5}$  &  0.014  &   100         & 0.0022  &100           & 0.00017 & 100\\
0.5  &              $10^{-5}$  &  1.429  & 100         & 0.227     & 100           & 0.029  &100\\
1    &              $10^{-5}$  &  9.608  & 8--31.2$\,^a$ & 1.682   & 100        & 0.316    & 100\\
2    &              $10^{-5}$  &  66.01  &  2.57--4.5& 13.84       & 12.3--21.7& 4.747    & 35.8--63.2$\,^a$\\
3    &              $10^{-5}$  &  211.5  & 2.65--5.15& 52.86       & 10.6--20.6& 23.37    & 23.9--46.6\\
5    &              $10^{-5}$  &  956.2  & 0.65--1.3& 299          & 2--4.1& 169.6       & 3.65--7.3\\
\hline
0.1  &              $10^{-6}$  &  0.047  & 100         & 0.014      &  100          & 0.0023  &100\\
0.5  &              $10^{-6}$  &  6.275  & 62.15--100$\,^a$ & 1.776 &  100          & 0.0246  & 100\\
1    &              $10^{-6}$  &  44.6   & 3.87--6.72  & 12.8       & 13.5--23.4& 2.313       & 74.8--100$\,^a$\\
2    &              $10^{-6}$  &  319    &  0.53--0.94& 96.14       & 1.77--3.12& 24.95       & 6.8--12\\
3    &              $10^{-6}$  &  1030   &0.54--1   & 326.2         & 1.7--3.34& 107.8        & 5.2--10.1\\
5    &              $10^{-6}$  &  4653   & 0.13--0.35& 1600         & 0.38--0.78& 690.8       & 0.9--1.8\\
\hline
0.1  &              $10^{-7}$  &  0.150  & 100              & 0.052   & 100        & 0.014 & 100 \\
0.5  &              $10^{-7}$  &  28.2   & 13.8--33   & 8.673         & 45--100$\,^a$      & 100 \\
1    &              $10^{-7}$  &  210.0  & 0.82--1.42 & 65.75         & 2.63--4.56       & 16.8  & 10.3--55.35\\
2    &              $10^{-7}$  &  1565   & 0.1--0.2   & 500.3         & 0.34--0.6& 153.2         &1.2--1.9\\
3    &              $10^{-7}$  &  5270   & 0.1--0.3   & 1680          & 0.47--0.92&584.1         &0.95--1.86\\
5    &              $10^{-7}$  &  28620  & 0.02--0.25 & 8117          &  0.09--0.2& 3250         & 0.24--0.49\\
\hline
0.1  &              $10^{-8}$  &  0.392  & 100       & 0.183   &  100       & 0.054 & 100 \\
0.5  &              $10^{-8}$  &  116.7  & 9.4--20.4 & 41.95   &  26.2--56.7& 10.75 & 100\\
1    &              $10^{-8}$  &  1002   &    1.73--4& 332.6   & 5.2--12.1  & 93.2  & 1.85--3.2\\
2    &              $10^{-8}$  &  9934   & 1.93--4   & 2670    & 0.23--0.43 & 826.1 & 0.72--1.39\\
3    &              $10^{-8}$  &  56640  & 0.19--0.46& 9797    & 1.1--2.65  & 3066  & 0.26--0.5\\
5    &              $10^{-8}$  &  105816 & 0.03--0.07& 103725  & 0.03--0.072& 17810 & 0.23--0.42\\
\hline
\end{tabular}
\end{center}
$^a)$ Protoplanets that may lose their remaining hydrogen remnants
during the rest of their lifetime after the XUV activity saturation phase ends
and the solar/stellar XUV flux begins to decrease (see Sect. 2) to $\leq$
the present time solar value at 1 AU \cite{Erkaev2013}.
\normalsize
\label{t_mnras}
\end{table}

As it was shown by \cite{Erkaev2013, Kislyakova2013}, if an exoplanet located in the HZ does not lose the dense primordial envelope during the first 90 Myr years of star's maximal activity (this period is considered by \cite{Lammer2014}), it is unrealistic that this additional volatiles will be lost during the next 4.5 Gyr. This can lead to the formation of a mini-Neptune exoplanet. Only the close exoplanets can lose the envelopes completely \cite{Leitzinger2011}.

%conclusions
From the results of our study we find that the nebula properties, protoplanetary growth-time, planetary mass, size and the host stars radiation environment set the initial conditions for planets that can evolve as the Earth-like class I habitats \cite{Lammer2009}.

 We found that protoplanets with core masses that are $\leq$1M$_{\rm \oplus}$ can lose their captured hydrogen envelopes during the active XUV saturation phase of their host stars, while rocky cores within the so-called super-Earth domain most likely can not get rid of
their nebula captured hydrogen envelopes during their whole lifetime. Our results are in agreement with the suggestion that Solar System terrestrial planets, such as Mercury, Venus, Earth and Mars, lost their nebula-based protoatmospheres during the XUV activity saturation phase of the young Sun. We also conclude that several recently discovered low density super-Earths with known radius and mass even at closer orbital distances could not get rid of their hydrogen envelopes. Furthermore, our results indicate that one should expect many super-Earths to be discovered in the near future inside habitable zones with hydrogen dominated atmospheres.

\section{Ion Pickup}
\label{sec:nth}

Besides thermal escape, various nonthermal escape mechanisms also contribute to the total mass loss and should be estimated. Nonthermal escape processes can be separated in ion escape and photochemical and kinetic processes that accelerate atoms beyond escape energy. Ions can escape from an upper atmosphere if the exosphere is not protected by a strong magnetic field and stretches above the magnetopause. In such a case, exospheric neutral atoms can interact with the stellar plasma  environment. Also, planetary ions can be detached from an ionopause by plasma instabilities in the form of ionospheric clouds \cite{Terada2002}.

In this section we present our estimations of ion pickup loss, which is one of the most effective among them (according to analysis of data for Venus and Mars obtained by the ASPERA instruments on board Venus Express and Mars Express). We assume the exoplanets have no magnetic field to estimate the maximum possible amount of losses. In our model the ions are produces by charge-exchange with stellar protons, photoionization by stellar photons and electron impact ionization by electrons in the stellar wind.

Charge exchange reactions between stellar wind protons and neutral planetary particles consist of the transfer of an electron from a neutral atom to a proton leaving a cold atmospheric ion and an energetic neutral atom. This process is described by the following reaction: $H_{\rm sw}^+ + H_{\rm pl} \to H^+_{\rm pl} + H_{\rm ENA}$.

We used the DSMC method to model the stellar wind -- upper atmosphere interaction. The code starts at the inner boundary $R_0$ (Knudsen number 0.1 or 1 depending on the simulation). The details about the initial algorithm can be found in \cite{Holmstrom2008}, about the developed versions -- in \cite{Kislyakova2013, Kislyakova2014}. Here we repeat only the main things.

The code includes two species, neutral hydrogen atoms and protons. Following processes/forces can act at neutral $H$: 
\begin{enumerate}
  \item Collision with an UV photon, which can occur if the particle is outside of the planet's shadow. Leads to an acceleration of the hydrogen atom away from the star. A UV photon is absorbed by a neutral hydrogen atom, leading to a radial velocity change, and then consequently reradiated in a random direction. The UV collision rate is velocity dependent (depends also on the star Ly-$\alpha$ flux and distance to the planet).
  \item Photoionization by a stellar photon or a stellar wind electron impact ionization. 
  \item Charge exchange between neutral hydrogen atoms and stellar wind protons. If a hydrogen atom is outside the planetary obstacle (magnetopause or ionopause) it can charge exchange with a stellar wind proton, producing an energetic neutral atom (ENA) and an initially cold planetary ion. The charge exchange cross-section is taken to be equal to $2 \times 10^{-19}$~m$^2$ \cite{Lindsay2005}.
  \item Elastic collision with another hydrogen atom. Here the collision cross section was taken to be $10^{-21}$~m$^2$ \cite{Izmodenov2000}.
\end{enumerate}

The coordinate system is centered at the center of the planet with the $x$-axis pointing towards the center of mass of the system, the $y$-axis pointing in the direction opposite to the planetary motion, and the $z$-axis pointing parallel to the vector $\Omega$ representing the orbital angular velocity of the planet around the central star. $M_{\rm st}$ is the mass of the planet's host star. The outer boundary of the simulation domain is the box
$x_{\rm min} \leq x \leq x_{\rm max}$,~
$y_{\rm min} \leq y \leq y_{\rm max}$, and
$z_{\rm min} \leq z \leq z_{\rm max}$. The inner boundary is a sphere of radius $R_0$.

Tidal potential, Coriolis and centrifugal forces, as well as the gravitation of the star and planet, acting on a hydrogen neutral atom are included in the following way (after \cite{Chandrasekhar1963}):

\begin{eqnarray}
  \frac{d v_{\rm i}}{dt}=\frac{\partial}{\partial x_{\rm i}} \Bigg[ \frac{1}{2}\Omega^2 \left(x_1^2+x_2^2 \right)+ \mu\left(x_1^2-\frac{1}{2}x_2^2 -\frac{1}{2}x_3^2 \right)+ \left(\frac{G M_{\rm st}}{R^2} - \frac{M_{\rm st} R}{M_{\rm pl} + M_{\rm st}}\Omega^2 \right)x_1 \Bigg]+\nonumber\\
+ 2\Omega \epsilon_{\rm il3} v_{\rm l}~~~~~~~~~
  \label{e_SCG}
\end{eqnarray}

Here $x_1 = x$, $x_2 = y$, $x_3 = z$, $v_{\rm i}$ are the components of the velocity vector of a particle, $G$ is Newton's gravitational constant, $R$ is the distance between the centers of mass, $\epsilon_{\rm ilk}$ is the Levi-Civita symbol, and $\mu$ is given by $\mu = G M_{\rm st}/R^3$. The first term in the right-hand side of Equation \ref{e_SCG} represents the centrifugal force, the second is the tidal-generating potential, the third corresponds to the gravitation of the planet's host star and the planet, while the last term stands for the Coriolis force. The self-gravitational potential of the particles is neglected.

Charge exchange takes place outside of an obstacle that corresponds to the magnetopause or ionopause boundary, which we assume is a surface described by:
\begin{equation}
	x = R_{\rm s} \left(1 - \frac{y^2 + z^2}{R_{\rm t}^2} \right)
\label{e_obs}
\end{equation}

Intensity of the stellar wind atmosphere erosion and, consequently, influence on the planetary evolution is different for different aged stars and thus is strongly connected with the evolution of the star. It was discussed in \cite{Erkaev2013} and \cite{Kislyakova2013} that the stars of late spectral classes (red K and mostly M dwarfs) can keep high levels of the XUV radiation. The stellar wind evolution is more controversial \cite{Wood2005}, but higher XUV and X-ray heating intensity the nonthermal erosion of the atmosphere even if one considers the interaction with the constant stellar wind.

\begin{figure}
\includegraphics[width=1.0\columnwidth]{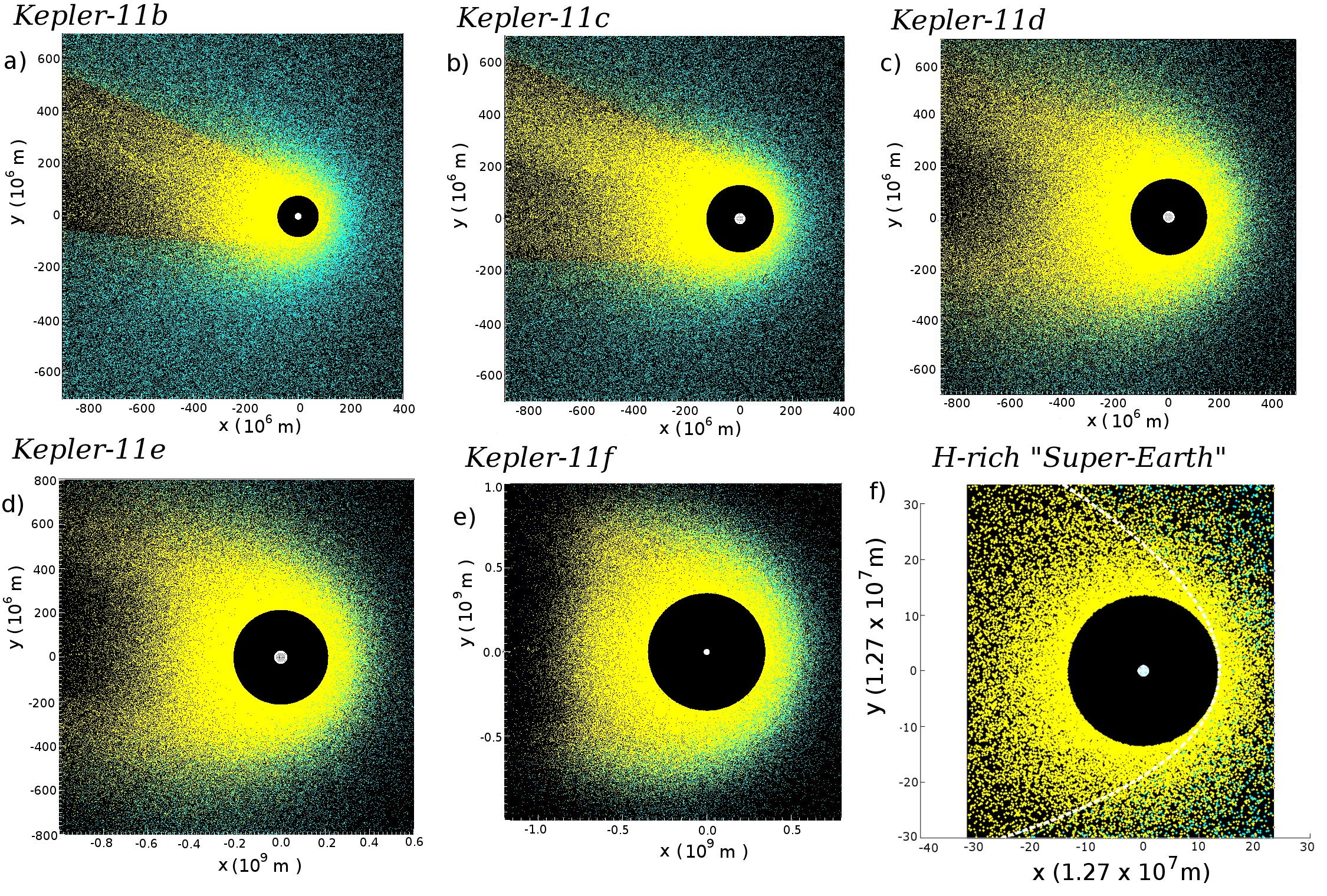}
\caption{a--e: Slices of modeled 3D atomic hydrogen coronae around the five Kepler-11 super-Earths for  $-10^7 \le z \le 10^7$~m and heating efficiency $\eta$ = 40\%. Yellow and green dots correspond to neutral hydrogen atoms and hydrogen ions, including stellar wind protons, respectively. The white dot in the center represents the planet. The black empty area around the planet corresponds to the XUV heated, hydrodynamically expanding thermosphere up to the height $R_0$ where $Kn$=0.1 \cite{Kislyakova2014}. f: Modeled atomic hydrogen coronae and stellar wind plasma interaction around a super-Earth hydrogen-rich planet inside an M star HZ at 0.24 AU. The XUV flux is 50 times higher than that of the present Sun, $\eta$=15\%; the dashed line denotes the planetary obstacle.}
\label{f_clouds}
\end{figure}

Fig.\ref{f_clouds} presents the results of the DSMC modeling of hydrogen coronae around five Kepler-11 super-Earths \cite{Kislyakova2014} and around a ``test'' super-Earth with $M = 10 M_\oplus$, $R = 2 R_\oplus$ located in the habitable zone of an M dwarf \cite{Kislyakova2013}. As one may see, in each case a huge hydrogen corona is formed composed of neutral hydrogen of planetary origin, ENAs and $H$ atoms accelerated ny the radiation pressure. Radiation pressure effects are of most importance in the vicinity of the host star (Kepler-11b and -c). This type of acceleration was considered by \cite{Bourrier2013, Lecavelier2004} in application to the Hot Jupiters, where these effects of even higher importance are.

In all cases we considered non-magnetized planets, where the magnetic obstacle defined by equation \ref{e_obs} is located very close to the planets. Possible magnetic moments of exoplanets are discussed, for example, in \cite{Lammer2009b} and for Kepler-11b--f are believed to be rather weak.

%-----------------------------------------------------------------
%-----------------------------------------------------------------

%---------------------------------------------------------------------
\begin{table}
\renewcommand{\baselinestretch}{1}
\caption{Ion pickup loss rates and thermal escape rates for Kepler-11 planets for heating efficiencies of 15 \% and 40 \%. The values are given in [g$\cdot$s$^{-1}$]. Thermal escape rates are taken from Table 3 in \cite{Lammer2013b}.}
\begin{center}
\begin{tabular}{ccccc}
\hline\hline
Exoplanet~~~&  $L_{\rm ion}$, $\eta=$15\%~~~& $L_{\rm th}$, $\eta=$15\%~~~& $L_{\rm ion}$, $\eta=$40\%~~~&  $L_{\rm th}$, $\eta=$40\%~~~\\\hline
Kepler-11b &   $\sim 1.17 \times 10^{7}$ & $\sim 1.15 \times 10^{8}$  &  $\sim 1.3 \times 10^{7}$ & $\sim 2.0 \times 10^{8}$ \\

Kepler-11c & $\sim 1.07 \times 10^{7}$ &  $\sim 4.0 \times 10^{7}$  &  $\sim 1.37 \times 10^{7}$ &  $\sim 1.3 \times 10^{8}$ \\

Kepler-11d & $\sim 1.47 \times 10^{7}$ &  $\sim 1.0 \times 10^{8}$  &  $\sim 2.33 \times 10^{7}$ &  $\sim 2.5 \times 10^{8}$ \\

Kepler-11e & $\sim 1.84 \times 10^{7}$ &  $\sim 1.1 \times 10^{8}$  &  $\sim 3.34 \times 10^{7}$ &  $\sim 2.5 \times 10^{8}$ \\

Kepler-11f & $\sim 6.0 \times 10^{7}$  &  $\sim 4.0 \times 10^{8}$  &  $\sim 6.8 \times 10^{7}$  &  $\sim 4.5 \times 10^{8}$ \\
\hline
\end{tabular}
\end{center}
\normalsize
\label{t_AA}
\end{table}
%---------------------------------------------------------------------

Table \ref{t_AA} illustrates average ion production rates, $L_{\rm ion}$, depending on coronal conditions related to heating efficiency $\eta$.

We found  that in the Kepler-11 system the non-thermal loss rates are approximately one order of magnitude smaller than the thermal losses estimated for the same planets by \cite{Lammer2013a} (for detailed comparison and discussion see \cite{Kislyakova2014}). This ratio is in good agreement with results obtained by \cite{Kislyakova2013} and \cite{Erkaev2013} for an Earth-type planet and a super-Earth in the habitable zone of a GJ436-like M-type host star. 

\section{Conclusions }
\label{sec:concl}

In this chapter we considered escape rates and their influence on atmospheric evolution of medium-size exoplanets. It was shown, that the size, mass and distance to the star together with the star-type define in many aspects, if the planet may evolve into a terrestrial-type exoplanets. The amount of initially accumulated gases, i.e. primordial nebula properties, plays a big role as well \cite{Lammer2014, Ikoma2006}. In general, close-in exoplanets lose their primordial envelopes and secondary atmospheres more easily, they easily reach the blow-off state and experience it longer \cite{Lammer2013b} and lose their atmospheres also due to Roche-lobe overflow \cite{Erkaev2007}. In the HZ of the parent star the planetary evolution strongly depends on the planet's mass. According to results of \cite{Lammer2014} Mars-sized bodies and planets with masses up to $<5 M_\oplus$ can lose significant percentage or even a whole of their atmospheres. On the other hand, so-called super-Earths can experience difficulties in losing their dense primordial envelopes composed mostly of light gases like hydrogen \cite{Erkaev2013, Lammer2013b}.

Nonthermal ion pickup escape contributes to the total atmospheric losses as well, but although they are of most importance of small-sized planets like Mars \cite{Lundin2011}, they make up only several percent of total thermal and nonthermal losses and can not change the evolutionary scenarios significantly \cite{Kislyakova2013, Kislyakova2014}. 

Atmospheric evolution of an exoplanet is also deeply connected with the evolution of its host star. M dwarfs live longer and develop slower in comparison to Sun-type stars, which means that they stay longer in the highest activity saturation phase. If an exoplanet orbits an M dwarf, it experiences severe stellar conditions (high levels of X-rays and EUV radiation, stronger stellar wind) much longer, leading to additional losses in comparison to an exoplanet of the same type orbiting a G dwarf \cite{Erkaev2013, Kislyakova2013}.

Summarize the above mentioned facts, one may conclude that forming of a terrestrial-type nitrogen atmosphere requires several restrictions on an exoplanet and its host star. The star should be a long-lived object, which probably excludes the very early spectral classes; on the other hand, it should stay not too long in the highest activity stage, when the planetary atmospheres undergo strong erosion. The latter makes the fate of the exoplanets in M dwarf systems controversial, however, it is not excluded that the planets may evolve as a terrestrial-type body also orbiting these stars. As for exoplanet itself, it has to be located in the right distance from its star, gain enough mass and volatiles during the formation. The volatile envelope should be thick enough to protect the atmospheres from the high stellar activity during the early age, but thin enough to be lost. Otherwise the exoplanet would evolve in a mini-Neptune containing up to several percents of its weight in the light gases.

It is worth to mention that Earth satisfies all above mentioned restrictions, which can be not a coincidence.

%\begin{acknowledgment}
\verb|Acknowledgements.|
K.G. Kislyakova, and H. Lammer acknowledge support by the FWF NFN project S116 "Pathways to Habitability: From Disks to Active Stars, Planets and Life," and the related FWF NFN subprojects, S116 604-N16 "Radiation \& Wind Evolution from T Tauri Phase to ZAMS and Beyond," S116 606-N16 "Magnetospheric Electrodynamics of Exoplanets", S11 6607-N16 "Particle/Radiative Interactions with Upper Atmospheres of Planetary Bodies Under Extreme Stellar Conditions." 
%\end{acknowledgment}

%\section{Acknowledgments}
%Work at the Institute of Nuclear Physics,
%Moscow State University was supported by the RFBR Grants
%11-05-00894-, 12-05-00219-, and 12-02-92600-_, by the EU
%FP7 projects EUROPLANET/JRA3 and IMPEX, and by the Ministry of
%Education and Science of the Russian Federation Grant No 07.514.11.4020


\begin{thebibliography}{}

\bibitem{Lundin2011}
{Lundin}, R. (2011)
Ion Acceleration and Outflow from Mars and Venus: An Overview, SSR, 162, 309.

\bibitem{Vidal-Madjar2003}
{Vidal-Madjar}, A. and {Lecavelier des Etangs}, A. and {D{\'e}sert}, J.-M. and 	{Ballester}, G.~E. and {Ferlet}, R. and {H{\'e}brard}, G. and 	{Mayor}, M. (2003)
An extended upper atmosphere around the extrasolar planet HD209458b, Nature, 422, 143.

\bibitem{Lecavelier2010}
{Lecavelier Des Etangs}, A. and {Ehrenreich}, D. and {Vidal-Madjar}, A. and {Ballester}, G.~E. and {D{\'e}sert}, J.-M. and {Ferlet}, R. and {H{\'e}brard}, G. and {Sing}, D.~K. and {Tchakoumegni}, K.-O. and {Udry}, S. (2010)
Evaporation of the planet HD 189733b observed in H I Lyman-{$\alpha$}. A\&A, 514, A72.

\bibitem{Bourrier2013}
{Bourrier}, V. and {Lecavelier des Etangs}, A. (2013)
3D model of hydrogen atmospheric escape from HD 209458b and HD 189733b: radiative blow-out and stellar wind interactions, A\&A, 557, A124.

\bibitem{Broeg2009}
{Broeg}, C.~H. (2009)
The full set of gas giant structures I: On the origin of planetary masses and the planetary initial mass function. Icarus, 204, 15.

\bibitem{Ikoma2006}
{Ikoma}, M. and {Genda}, H. (2006)
Constraints on the Mass of a Habitable Planet with Water of Nebular Origin, ApJ, 648, 696.

\bibitem{Lammer2013a}
{Lammer}, H. and {Kislyakova}, K.~G. and {G{\"u}del}, M. and {Holmstr{\"o}m}, M. and {Erkaev}, N.~V. and {Odert}, P. and {Khodachenko}, M.~L. (2013)
Stability of Earth-Like N$_{2}$ Atmospheres: Implications for Habitability. In: The Early Evolution of the Atmospheres of Terrestrial Planets, Astrophysics and Space Science Proceedings (Springer, New York), 33-52.

\bibitem{Lammer2014}
{Lammer}, H. and {Erkaev}, N.~V. and {Odert}, P. and {Kislyakova}, K.~G. and {Leitzinger}, M. and {Khodachenko}, M.~L. (2014)
Loss of nebula-captured hydrogen envelopes from "sub"- to "super-Earths" in the habitable zone of Sun-like stars. MNRAS, accepted.

\bibitem{Elkins-Tanton2008}
{Elkins-Tanton}, L.~T. and {Seager}, S. (2008)
Ranges of Atmospheric Mass and Composition of super-Earth Exoplanets. ApJ, 685, 1237.

\bibitem{Tian2008a}
{Tian}, F. and {Kasting}, J.~F. and {Liu}, H.-L. and {Roble}, R.~G. (2008)
Hydrodynamic planetary thermosphere model: 1. Response of the Earth's thermosphere to extreme solar EUV conditions and the significance of adiabatic cooling. JGR (Planets), 113, 5008.

\bibitem{Tian2008b}
{Tian}, F. and {Solomon}, S.~C. and {Quian}, L. and {Lei}, J. (2008)
Hydrodynamic planetary thermosphere model: 1. Coupling of an electron transport/energy deposition model. JGR (Planets), 113, 7005.

\bibitem{Lichtenegger2010}
{Lichtenegger}, H.~I.~M. and {Lammer}, H. and {Grie{\ss}meier}, J.-M. and {Kulikov}, Y.~N. and {von Paris}, P. and {Hausleitner}, W. and {Krauss}, S. and {Rauer}, H. (2010). Aeronomical evidence for higher CO$_{2}$ levels during Earth's Hadean epoch, Icarus, 210, 1.

\bibitem{Hunten1987}
{Hunten}, D.~M. and {Pepin}, R.~O. and {Walker}, J.~C.~G. (1987)
Mass fractionation in hydrodynamic escape. Icarus, 69, 532-549.

\bibitem{Kulikov2006}
{Kulikov}, Yu.~N., and {Lammer}, H., and {Lichtenegger}, H.I.M., and {Terada}, N., and {Ribas}, E., and {Kolb}, C., and {Langmayr}, D., and {Lundin}, R., and {Guinan}, E.F., and {Barabash}, S., and {Biernat}, H.B. (2006)
Atmospheric and water loss from early Venus, PSS, 54, 1425.

\bibitem{Erkaev2007}
{Erkaev}, N.~V. and {Kulikov}, Y.~N. and {Lammer}, H. and {Selsis}, F. and {Langmayr}, D. and {Jaritz}, G.~F. and {Biernat}, H.~K. (2007)
Roche lobe effects on the atmospheric loss from "Hot Jupiters". A\&A, 472, 329.

\bibitem{Lecavelier2004}
{Lecavelier des Etangs}, A. and {Vidal-Madjar}, A. and {McConnell}, J.~C. and {H{\'e}brard}, G. (2004).
Atmospheric escape from hot Jupiters. A\&A, 418, L1-L4.

\bibitem{Leitzinger2011}
{Leitzinger}, M. and {Odert}, P. and {Kulikov}, Y.~N. and {Lammer}, H. and {Wuchterl}, G. and {Penz}, T. and {Guarcello}, M.~G. and {Micela}, G. and 	{Khodachenko}, M.~L. and {Weingrill}, J. and {Hanslmeier}, A. and {Biernat}, H.~K. and {Schneider}, J. (2011).
Could CoRoT-7b and Kepler-10b be remnants of evaporated gas or ice giants? PSS, 59, 1472.

\bibitem{Lammer2009}
{Lammer}, H. and {Bredeh{\"o}ft}, J.~H. and {Coustenis}, A. and {Khodachenko}, M.~L. and {Kaltenegger}, L. and {Grasset}, O. and {Prieur}, D. and {Raulin}, F. and {Ehrenfreund}, P. and {Yamauchi}, M. and {Wahlund}, J.-E. and {Grie{\ss}meier}, J.-M. and {Stangl}, G. and 	{Cockell}, C.~S. and {Kulikov}, Y.~N. and {Grenfell}, J.~L. and 	{Rauer}, H. (2009)
 What makes a planet habitable? The A\&Astrophysics Review, 17, 181-249.

\bibitem{Erkaev2013}
{Erkaev}, N.~V. and {Lammer}, H. and {Odert}, P. and {Kulikov}, Y.~N. and 
	{Kislyakova}, K.~G. and {Khodachenko}, M.~L. and {G{\"u}del}, M. and 
	{Hanslmeier}, A. and {Biernat}, H. (2013)
XUV exposed non-hydrostatic hydrogen-rich upper atmospheres of terrestrial planets. Part I: Atmospheric expansion and thermal escape. Astrobiology, 11, 1-19.

\bibitem{Guedel2007}
{G{\"u}del}, M. (2007).
The Sun in Time: Activity and Environment, Living Reviews in Solar Physics, 4, 3.

\bibitem{Khodachenko2007}
{Khodachenko}, M.~L. and {Ribas}, I. and {Lammer}, H. and {Grie{\ss}meier}, J.-M. and {Leitner}, M. and {Selsis}, F. and {Eiroa}, C. and {Hanslmeier}, A. and 
	{Biernat}, H.~K. and {Farrugia}, C.~J. and {Rucker}, H.~O. (2007)
Coronal Mass Ejection (CME) Activity of Low Mass M Stars as An Important Factor for The Habitability of Terrestrial Exoplanets. I. CME Impact on Expected Magnetospheres of Earth-Like Exoplanets in Close-In Habitable Zones. Astrobiology, 7, 167-184.

\bibitem{Kislyakova2013}
{Kislyakova}, K.~G. and {Lammer}, H. and {Holmstr{\"o}m}, M. and 
	{Panchenko}, M. and {Odert}, P. and {Erkaev}, N.~V. and {Leitzinger}, M. and 
	{Khodachenko}, M.~L. and {Kulikov}, Y.~N. and {G{\"u}del}, M. and 
	{Hanslmeier}, A. (2013)
XUV exposed, non-hydrostatic hydrogen-rich upper atmospheres of terrestrial planets II: Hydrogen coronae and ion escape. Astrobiology, 11, 1030-1048.

\bibitem{Lammer2013b}
{Lammer}, H. and {Erkaev}, N.~V. and {Odert}, P. and {Kislyakova}, K.~G. and {Leitzinger}, M. and {Khodachenko}, M.~L. (2013)
Probing the blow-off criteria of hydrogen-rich "super-Earths". MNRAS, 430, 1247-1256.

\bibitem{Terada2002}
{Terada}, N. and {Machida}, S. and {Shinagawa}, H. (2002)
Global hybrid simulation of the Kelvin-Helmholtz instability at the Venus ionopause. JGR (Space Physics), 107, 1471.

\bibitem{Holmstrom2008}
{Holmstr{\"o}m}, M. and {Ekenb{\"a}ck}, A. and {Selsis}, F. and {Penz}, T. and {Lammer}, H. and {Wurz}, P. (2008)
Energetic neutral atoms as the explanation for the high-velocity hydrogen around HD 209458b. Nature, 451, 970-972.

\bibitem{Kislyakova2014}
{Kislyakova}, K.~G. and {Johnstone}, C.~P. and {Odert}, P. and {Erkaev}, E.~V. and {Lammer}, H. and {L\"{u}ftinger}, T., and {Holmstr\"{o}m}. M., and {Khodachenko}, M.~L., and {G\"{u}del}, M. (2014)
Stellar wind interaction and pick-up ion escape of the Kepler-11 "super-Earths". A\&A, accepted.

\bibitem{Lindsay2005}
{Lindsay}, B.~G. and {Stebbings}, R.~F. (2005)
Charge transfer cross sections for energetic neutral atom data analysis. JGR, 110, 12213.

\bibitem{Izmodenov2000}
{Izmodenov}, V.~V. and {Malama}, Y.~G. and {Kalinin}, A.~P. and  {Gruntman}, M. and {Lallement}, R. and {Rodionova}, I.~P. (2000)
Hot Neutral H in the Heliosphere: Elastic H-H, H-p Collisions. APSS, 274, 71-76.

\bibitem{Chandrasekhar1963}
{Chandrasekhar}, S. (1963)
The Equilibrium and the Stability of the Roche Ellipsoids. ApJ, 138, 1182

\bibitem{Wood2005}
{Wood}, B.~E. and {M{\"u}ller}, H.-R. and {Zank}, G.~P. and {Linsky}, J.~L. and {Redfield}, S. (2005)
New Mass-Loss Measurements from Astrospheric Ly{$\alpha$} Absorption. ApJL, 628, L143.

\bibitem{Lammer2009b},
{Lammer}, H. and {Odert}, P. and {Leitzinger}, M. and {Khodachenko}, M.~L. and 	{Panchenko}, M. and {Kulikov}, Y.~N. and {Zhang}, T.~L. and {Lichtenegger}, H.~I.~M. and {Erkaev}, N.~V. and {Wuchterl}, G. and {Micela}, G. and {Penz}, T. and {Biernat}, H.~K. and {Weingrill}, J. and {Steller}, M. and {Ottacher}, H. and {Hasiba}, J. and {Hanslmeier}, A. (2009),
Determining the mass loss limit for close-in exoplanets: what can we learn from transit observations?, A\&A, 506, 399-410.


\end{thebibliography}
\end{document}